\documentclass[%
 aip,
 amsmath,amssymb,
 10pt,
 twocolumn,
 longbibliography
]{revtex4}  

\usepackage{graphicx}
\usepackage{dcolumn}
\usepackage{bm}
\usepackage[utf8]{inputenc}
\usepackage[T1]{fontenc}
\usepackage{mathptmx}
\usepackage{color}
\usepackage{chemfig}
\usepackage{textcomp}

\begin{document}
 
 \def \bmath #1 {{\hbox{\boldmath{$#1$}\unboldmath}}}

 \title{\bf Spin-orbit transitions in the N$^+$($^3P_{J_A}$) + H$_2$ $\rightarrow$ NH$^+$($X^2\Pi$, $^4\Sigma^-$)+ H($^2S$) reaction,
   using adiabatic and mixed quantum-adiabatic statistical approaches
 }
 
 \author{Susana G\'omez-Carrasco}
\email{susana.gomez@usal.es}
 \affiliation{Facultad de Farmacia, Universidad de Salamanca,
      Campus Miguel de Unamuno, C. Lic. M{\'e}ndez Nieto, s/n, 37007-Salamanca, Spain}
 
\author{Daniel F\'elix-Gonz\'alez}
\affiliation{Unidad Asociada UAM-CSIC,
                       Departamento de Qu{\'\i}mica F{\'\i}sica Aplicada, Facultad de
                      Ciencias M-14, Universidad Aut\'onoma de Madrid, 28049, Madrid, Spain}
\author{Alfredo Aguado}
\affiliation{Unidad Asociada UAM-CSIC,
                       Departamento de Qu{\'\i}mica F{\'\i}sica Aplicada, Facultad de
                      Ciencias M-14, Universidad Aut\'onoma de Madrid, 28049, Madrid, Spain}

\author{Octavio Roncero}
\email{octavio.roncero@csic.es}
\affiliation{Instituto de F{\'\i}sica Fundamental (IFF-CSIC),  
                          C.S.I.C., 
                       Serrano 123, 28006 Madrid, Spain}
\begin{abstract}
  The cross section and rate constants for the title reaction are calculated for all the spin-orbit states
  of  N$^+$($^3P_{J_A}$) using two statistical approaches, one purely adiabatic and the other one mixing
  quantum capture for the entrance channel and  adiabatic treatment for the products channel. This is made
  by using a symmetry adapted basis set combining electronic (spin and orbital) and nuclear angular momenta
  in the reactants channel. To this aim, accurate {\it ab initio} calculations are performed separately
  for reactants and products. In the reactants channel,
  the three lowest electronic states (without spin-orbit couplings) have been diabatized, and the spin-orbit couplings
  have been introduced through a model localizing the spin-orbit interactions in  the N$^+$ atom, which yields accurate
  results as compared to {\it ab initio} calculations including spin-orbit couplings. For the products,
  eleven purely adiabatic spin-orbit states have been determined with {\it ab initio} calculations. The reactive
  rate constants thus obtained are in very good agreement with the available experimental data for several
  ortho-H$_2$ fractions, assuming a thermal initial distribution of spin-orbit states. The rate constants
  for selected spin-orbit $J_A$ states are obtained, to provide a proper validation of the spin-orbit effects
  to obtain  the experimental rate constants.
\end{abstract}

Submitted to Journal of Chemical Physics (2022), in press.
  \maketitle


\section{Introduction} \label{sec:introduction}

The formation of hydrides can be considered as the first step of chemistry in space
and determines the abundances of more complex molecules arising
in chemical networks from them. The study of the evolution of abundances of molecular species
allows the probe of physical conditions along the stellar evolution, from the parent molecular
cloud to the star system, passing through the intermediate stages such as cold and hot cores,
protoplanetary disk, etc. Among the most abundant elements, nitrogen plays a singular role, because
its more abundant forms are thought to be N$_2$ and atomic nitrogen, which are difficult to be observed
because they have no permanent dipole moment, specially
in cold cores. The abundance of nitrogen is then established by other molecules,
such as its hydrides NH$_n$, CN, HCN/HNC, N$_2$H$^+$, etc, requiring the construction
of increasingly more accurate chemical networks \cite{Wakelam-etal:10,LeGal-etal:13}.

Nitrogen hydrides are particularly interesting and ammonia is among the first polyatomic molecules
detected in the interstelar medium (ISM) \cite{Cheung-etal:68}. The ortho/para ratios observed
for NH$_2$ and NH$_3$ \cite{Persson-etal:10} and
their deuteration enrichment \cite{Hily-Blant-etal:13} serve as sensitive probes to check 
gas-phase chemistry models \cite{LeGal-etal:13}. In this regard, hydrides present a comparably
small number of reactions in the chemical networks.
In photodissociation regions (PDR), hydrides are normally formed from the atoms (neutral and/or cations, depending
on their ionization potential, as compared to atomic hydrogen)
  by successive addition of hydrogen atoms, followed by dissociative
  recombination with electrons in the case of cations.
  The ionization step in nitrogen in PDR is improbable difficult because its ionization energy
is larger than that of hydrogen, unlike most of other metal atoms, and the density of N$^+$ is therefore smaller.
Therefore other neutral reactions of N atoms with OH and CH  are alternative routes to form nitrogen hydrides \cite{LeGal-etal:13}.

The  rate constants involved in the first steps of the chemical networks have 
an enormous influence in the relative abundances, ortho/para ratios and deuteration fractions of many
of the nitrogen-bearing molecules. For these reasons
many experiments have been performed to study the following reaction
\cite{Marquette-etal:88,Sunderlin-Armentrout:94,Zymak-etal:13,Fanghanel:18}:
\begin{eqnarray}\label{H2yNp-reaction}
  N^+(^3P_{J_A}) + H_2(X^1\Sigma_g^+) \rightarrow  NH^+(^2\Pi,^4\Sigma^-)  + H(^2S)
\end{eqnarray}
These experiments are performed in different conditions, which raises questions about the 
reactivity associated to each fine structure state of N$^+$($^3P_{J_A}$), since the exact thermalization
conditions are not known.

Theoretical dynamical calculations have been performed on the ground adiabatic electronic state potential
\cite{Gonzalez-etal:86,Gonzalez-etal:89,Wilhelmsson-etal:92,Wilhelmsson-Nyman:92},
both classical \cite{Wilhelmsson-Nyman:92,Wilhelmsson-Nyman:92b}
and quantum \cite{Russell-Manolopoulos:99,Yang-etal:19} ones, without taking into account the fine structure of nitrogen.
These studies demonstrate that the reaction dynamics in the ground adiabatic state is mediated
by many long lived resonances due to the deep insertion well of the potential energy surface (PES).
{ These calculations suggest that the reaction proceeds statistically, but none of them describe
  any electronic transition among spin-orbit states}.

Several statistical simulations have been recently 
performed including the  fine structure \cite{Grozdanov-etal:15,Grozdanov-etal:16}. { However, in these statistical simulations
only long range interactions are included, within the assumption that only  the first 3 adiabatic fine-structure
states can react. However, the inclusion of transitions among the different spin-obit states in the entrance channel
may include important variations of the experimentally determined rate constants, specially at low temperature,
as it has been discussed by Zymak {\it et al.} \cite{Zymak-etal:13} and Fanghanel \cite{Fanghanel:18}.}

The main goal of this work is simulating the transitions between the fine structure N$^+$($^3P_{J_A}$) states, determining
the cross sections and rates for each of them individually.
Since the problem involves 9 spin-orbit states, some of them showing deep insertion wells,
complete quantum calculations are not feasible.
For this reason, in this work a detailed potential model is developed separately for reactants and products, all
based on accurate {\it ab initio}
calculations. In the  N$^+(^3P_{J_A})$ + H$_2$ reactants channel, a diabatic model is developed allowing
to include the  couplings among the spin-orbit states. In the
products channel, pure adiabatic spin-orbit potentials are calculated. These diabatic states
are used to build total electronic and angular basis sets allowing the study of the correlation
of angular momenta, electronic and nuclear, to properly describe spin-orbit transitions\cite{Jouvet-Beswick:87}.
These basis set functions are then used within an  adiabatic statistical (AS) approximation \cite{Quack-Troe:74}
and mixed description of the AS and a quantum statistical (QS)
\cite{Rackham-etal:01,Rackham-etal:03,Alexander-etal:04,Gonzalez-Lezana:07} (denoted by the acronym QAS).
The AS approximation has been recently applied to the study of many reactions and inelastic processes
for many systems and is widely used\cite{Konings-etal:21}.
A precedent of mixing quantum capture in the entrance channel and statistical approaches to describe the reaction
probability  has been proposed previously for four atom complex-forming reactions\cite{Mayneris-etal:06}.
In this work, the calculation of quantum capture probabilities is done with a time-independent
method  based on a renormalized
Numerov propagation scheme developed to this aim and presented in the Appendix \ref{appendix-zticc}.

This work is organized as follows.
A detailed {\it ab initio} study of the system will be described in section \ref{section-abinitio},
treating separately reactants, N$^+$($^3P_{J_A}$) + H$_2$($X^1\Sigma^+_g$), and products, H($^2S$)+ NH$^+$($^2\Pi, ^4\Sigma^-$)
channels, including spin-orbit couplings.
In the case of the reactants, a diabatic model is built  for the different N$^+$($^3P_{J_A}$) states, which is necessary 
to include the transitions among them. The PESs will be used to calculate the capture probabilities
needed in the quantum and adiabatic statistical methods, and described in section \ref{section-quantum-statistical},
paying special attention to the transitions among different fine structure states. Also cross sections
and rate coefficients for each individual state will be presented in section \ref{results}. Finally, in section \ref{section-conclusions},
some conclusion will be extracted.

\section{\label{section-abinitio} Potential energy surfaces}

\begin{figure}[t]
 \hspace*{-1.5cm}{
 \includegraphics[scale=0.4]{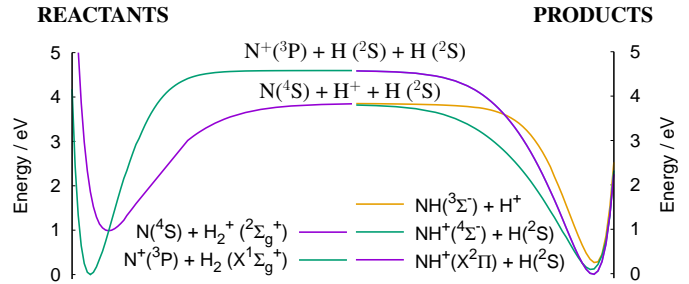}
 \vspace*{-4.0cm}
 \caption{\label{fragments}{\it  Asymptotic electronic states for reactants and products
}}
}
\end{figure}

An overall picture of the electronic states of reactants and products
of this system is displayed in  Fig. \ref{fragments}.
In the reactant region, the N($^4S$) + H$_2^+$(X$^2\Sigma_g^+$) channel
is located about 1 eV above the N$^+$($^3P_{J_A}$) + H$_2$($X^1\Sigma^+_g$) one so,
the former channel will not be populated at the energy range used in this work.
Regarding the product region, the lowest channels are
NH$^+$(X $^2\Pi$, a$^4\Sigma^-$) + H($^2S$) and NH($^3\Sigma^-$) + H$^+$ ones.

\subsection{ Ab initio calculations for the reactant channel}

\noindent

The $N^{+}(^3P_g) + H_2(X^1\Sigma_g^+)$ reactant channel has been calculated using a state-average complete active
space self-consistent field/multireference configuration interaction (SA-CASSCF/MRCI) method
with a VTZ-F12 explicitly correlated atomic basis set as implemented in the MOLPRO program\cite{MOLPRO-WIREs}.
Without taking spin-orbit coupling into account, three adiabatic electronic states, $^3\Pi$ and $^3\Sigma^{-}$,
correlate with the reactants in the C$_{\infty v}$ group of symmetry.
The calculations have been done in the $C_s$ point group of symmetry so that the state average multiconfigurational
wave function has included two $^3A''$ and one $^3A'$ states, with the molecule lying on the $y$-$z$ plane.
Subsequent MRCI energies have been obtained at the geometries described in the Supplementary Information (SI).
The ab initio points  have been interpolated using a
3D cubic spline method. Finally,  the long-range terms, charge-induced dipole and quadrupole \cite{Hirschfelder:67,Velilla-etal:08,Aguado-etal:21},
have been included for R> 15 a$_0$, using the following switching function of R centered at 20 a$_0$: 

\begin{eqnarray}
f(R) = \frac{1 - \tanh (0.5 \ \lbrack R - 20.0\rbrack)}{2}
\end{eqnarray}
The long-range terms included are described in detail in the Supplementary Information (SI), together
with some figures describing the main features of the PESs.

\begin{figure}[t]
\begin{center}
\includegraphics[width=9.cm]{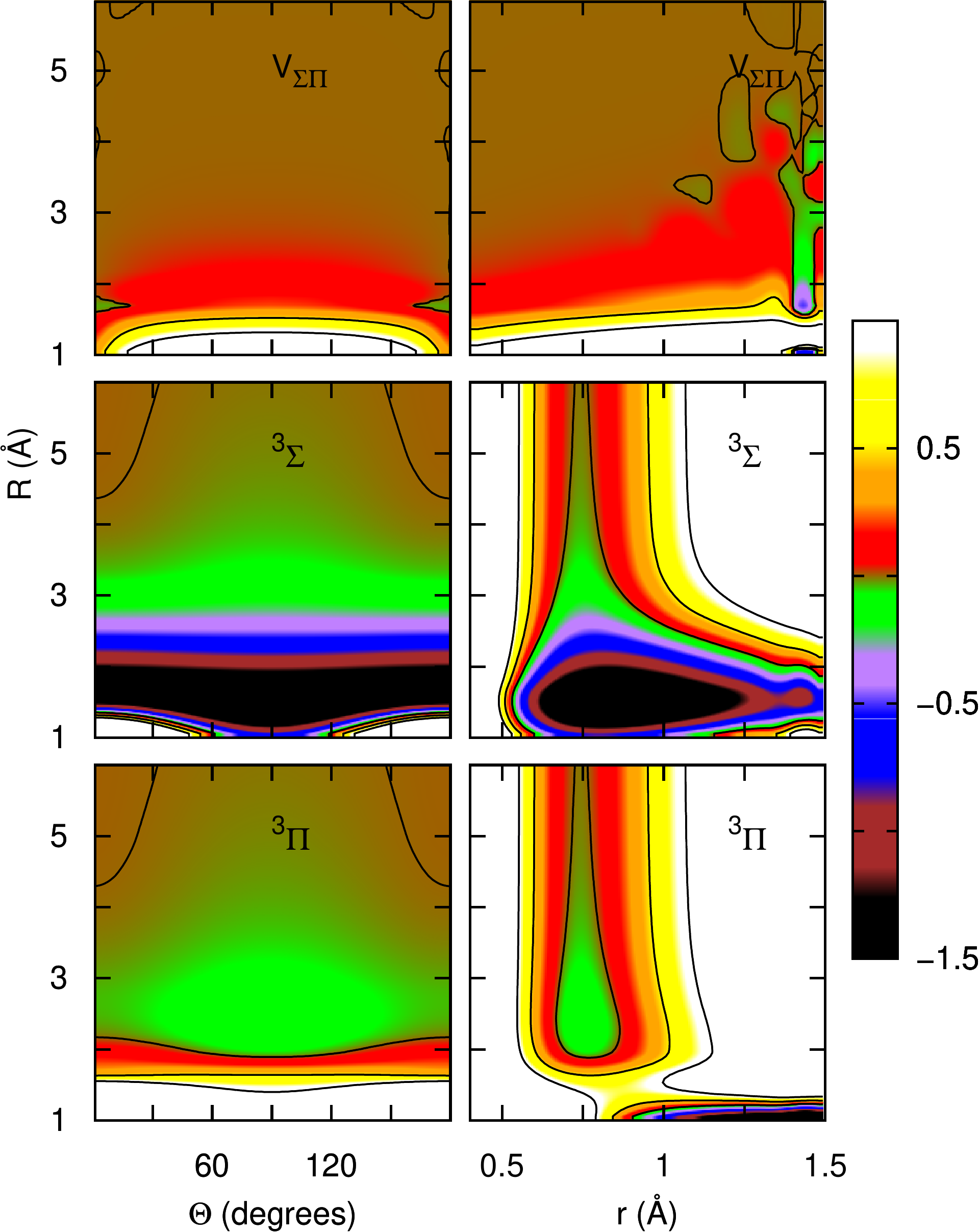}
  
 \caption{\label{diabatic-cuts}{\it Contour plots of the 
     PESs for the diabatic $^3\Sigma^-$,  $^3\Pi$ electronic
     states, and the $\Sigma-\Pi$ coupling, 
     obtained at the H$_2$ equilibrium distance,  $r$= 0.7 \AA\, as a function of the Jacobi distance, $R$,  and the angle $\gamma$ 
      (left panels)
      and at $\gamma$ = 90$^o$ as a function of $R$ and $r$ Jacobi distances.
      Energies are in eV, and the contour lines are at 0, 0.5 and 1 eV.
  }}
\end{center}
\end{figure}

Here we shall use a non-relativistic atomic basis set (hereafter called diabatic basis set)
$\vert L \Lambda S\Sigma\rangle$, where $L$ and $ S$ are the modula of the electronic orbital and spin
angular momenta of N$^+$, and $\Lambda$ and $\Sigma$ their projections, respectively, on the Jacobi body-fixed z-axis. In this basis,  
the non-relativistic electronic matrix takes the form \cite{Gomez-Carrasco-etal:06}
\begin{eqnarray}\label{diabatic-matrix}
  H = \left( \begin{array}{ccc}
               E_{-1}& V  & 0  \\
               V    & E_0 & V  \\
               0    & V   & E_1
  \end{array}\right)
  \quad {\rm with}\quad E_{-1}= E_1
\end{eqnarray}
whose eigenvalues correspond to the 1$^3A''$, 1$^3A'$ and 2$^3A''$
adiabatic electronic energies. The three unknown $E_1, E_0$ and $V$ in
Eq.~(\ref{diabatic-matrix})  can then be expressed in terms of
the {\it ab initio} energies as \cite{Gomez-Carrasco-etal:06}
\begin{eqnarray}\label{diabatic-energies}
  E_1 &=& E_{1^3A'} \nonumber\\
  E_0 &=& E_{1^3A''} +  E_{2^3A''} - E_{1^3A'}\\
  V &=& \sqrt{ {  (E_{1^3A''} -  E_{2^3A''})^2 -(E_1 -E_0)^2 \over 8}}.\nonumber
\end{eqnarray}
These diabatic energies are represented in Fig.~\ref{diabatic-cuts},
and the coupling $V$ in top panels reveal that the coupling between
the $\Sigma$ and $\Pi$ states become larger in the repulsive parts or
the adiabatic PESs, where the 1$^3A'$ and 2$^3A''$ differ the most.

\begin{figure}[t]
\begin{center}
\hspace*{-0.8cm}{
  \includegraphics[scale=0.45]{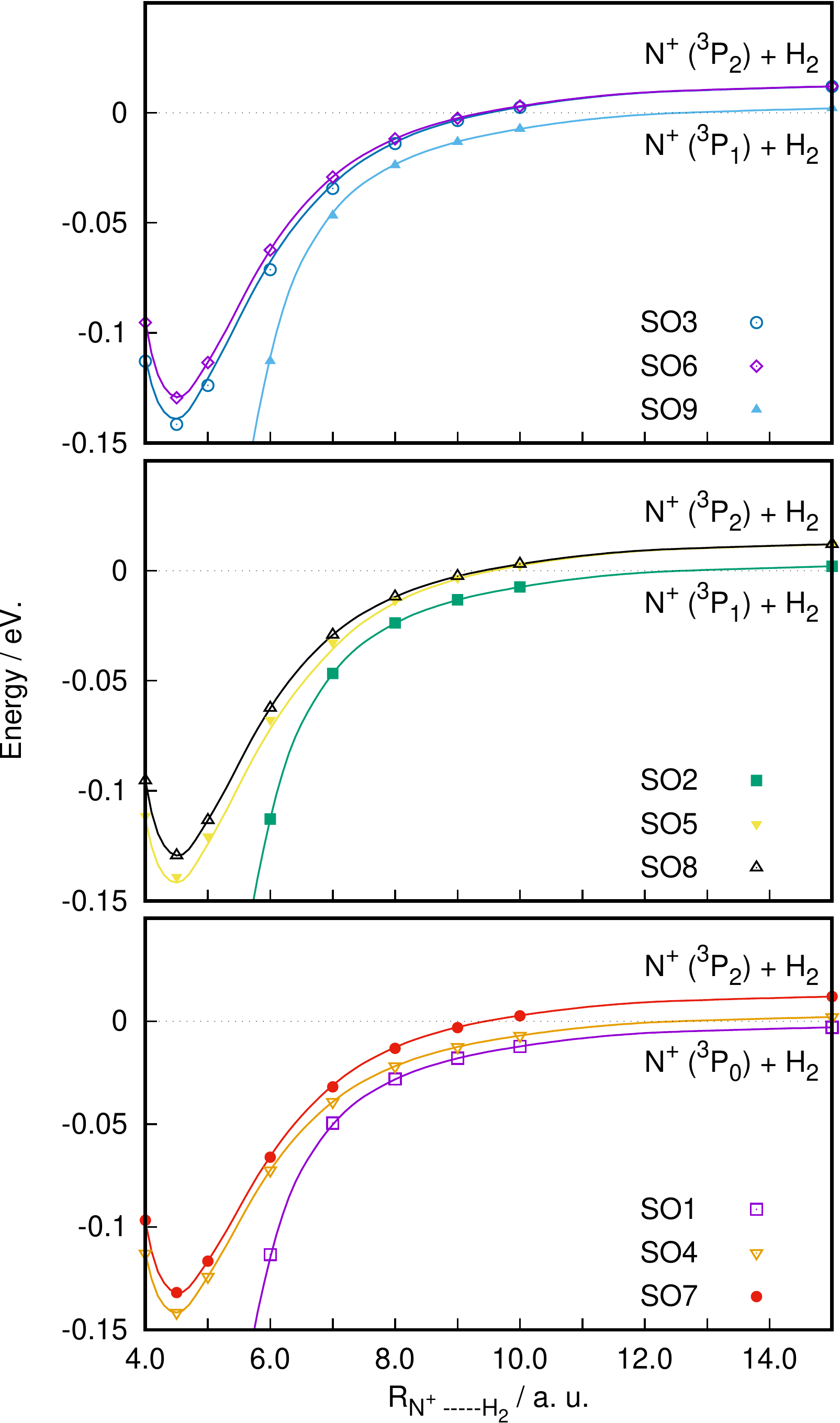}}
 \caption
 {\label{so-adiabatic-pes}{\it Energy profiles for the 9 spin-orbit electronic states correlating
     with N$^+$($^3P_{J_A=2, 1, 0}$)+H$_2$ as a function of the R Jacobi coordinate
     for $r$= 1.4 a.u. and $\gamma$=90 degrees. {The energy curves are distributed in three panels (1, 4 and 7 in the bottom panel,
     2, 5 and 8 in the middle panel and 3 ,6 and 9 in the top panel)
     to show more clearly the differences between {\it ab initio} (points) and the diabatic+atomic  spin-orbit model (lines),
     using the ab initio spin-orbit splittings.}
  }}
\end{center}
\end{figure}

The spin-orbit basis set, $\vert J_A \Omega_A\rangle$,
is expressed in terms of the diabatic
representation  $\vert L\Lambda S\Sigma\rangle$ defined above as
\begin{eqnarray}\label{spin-orbit-eigenfunctions}
  \vert J_A \Omega_A\rangle = \sum_{\Lambda,\Sigma} (-1)^{L-S+\Omega_A} \sqrt{2J_A+1)}
  \left({\scriptsize \begin{array}{ccc} L & S & J_A\\ \Lambda & \Sigma & -\Omega_A\end{array}}\right)
                                                            \vert L\Lambda S\Sigma\rangle, \nonumber\\
\end{eqnarray}
where  $\left({\scriptsize\begin{array}{c} \cdots \\ \cdots\end{array}}\right)$
are 3-j symbols.  Since H$_2$ is closed-shell, the total orbital (${\bf L}$) and spin  (${\bf S}$) electronic
angular momenta correspond  to atom N$^+(^3P_{J_A})$,
with ${\bf J}_A = {\bf L}+{\bf S}$, so that we shall consider
that H$_{SO}$ does not depend on the distance $R$,
and has eigenvectors $\vert J_A \Omega_A\rangle$, whose $E_{J_A}$ eigenvalues
are $(2J_A+1)$ degenerate. Following the treatment of Jouvet and Beswick \cite{Jouvet-Beswick:87},
summarized in the Supplementary Information for completeness,
the electronic Hamiltonian is expressed as $H= H_{SO} +  H_{el}$, with 
\begin{eqnarray}\label{non-relativistic-electronic-H}
  H_{el} = H_{el}^o + H_{el}^1 \quad{\rm with}\quad
 \lim_{\scriptsize R\rightarrow \infty} H_{el}^1 = 0,
\end{eqnarray}
$i.e.$, $H_{el}^1$ describes the  non-relativistic interaction
between H$_2$ and N$^+$, while $H_{el}^0$ describes  the two fragment
at infinity. The matrix elements of
$H_{el}^1$  are defined as (see SI and Ref \cite{Jouvet-Beswick:87})
\begin{eqnarray}\label{diabatic-spin-orbit-couplings}
  \left\langle  J_A \Omega_A  \left\vert H_{el}^1 \right\vert  J'_A \Omega'_A \right\rangle 
  &=&\sum_{\Lambda,\Lambda'} \sum_\Sigma  \sum_k (-1)^{\Omega_A-\Omega'_A} \\
\times \, V^K_{\Lambda\Lambda'} (r,R) && Y_{K\Lambda'-\Lambda}(\gamma,0)\sqrt{(2J_A+1)(2J'_A+1)}  \nonumber\\
  &\times&\left(\begin{array}{ccc} L & S & J_A\\ \Lambda & \Sigma & -\Omega_A\end{array}\right)
   \left(\begin{array}{ccc} L & S & J'_A\\ \Lambda' & \Sigma & -\Omega'_A\end{array}\right)
    ,\nonumber
\end{eqnarray}

In the $\vert J_A \Omega_A\rangle$ basis set, the atomic
spin-orbit Hamiltonian, $H_{SO}$, 
is diagonal. The experimental atomic spin-orbit splittings are 48.7 and 130.8 cm$^{-1}$ from NIST\cite{NIST}.

The 9 spin-orbit electronic states correlating with N$^+$($^3P_{J_A =0, 1, 2}$)+H$_2$
have been calculated at MRCI level using the Breit-Pauli operator.
At very long distances between N$^+$ and H$_2$, the {\it ab initio} calculations yield 40.2 cm$^{-1}$ and 120.6 cm$^{-1}$
for the energy of the N$^{+}(^3P_1)$ and N$^{+}(^3P_2)$ spin-orbit levels, respectively,
respect to the energy of the ground spin-orbit state N$^{+}(^3P_0)$.
These results are in good agreement with the experimental values.

In Fig. \ref{so-adiabatic-pes}, the adiabatic spin-orbit energies
obtained in the ab initio calculations are compared to those obtained
diagonalizing the $H_{el} + H_{SO}$ Hamiltonian (see SI for more information),
in which the spin-orbit term is considered to only affect N$^+(^3P_{J_A})$ subsystem, using the
ab initio spin-orbit splittings. The agreement is fairly good specially at long distances, and only
some discrepancies are found in the region of the bottom of the well.
This validates
the approximation of considering
the spin-orbit term only for the N$^+$ atom in the entrance channel.
Within this approximation, electronic transitions between all the spin-orbit states in the entrance
channel will be considered in the statistical calculations presented below, using
the experimental splittings.    \\

\subsection{Ab initio calculations for the product channels}

\begin{figure*}[t]
\begin{center}
\hspace*{-0.8cm}{
  \includegraphics[width=16cm]{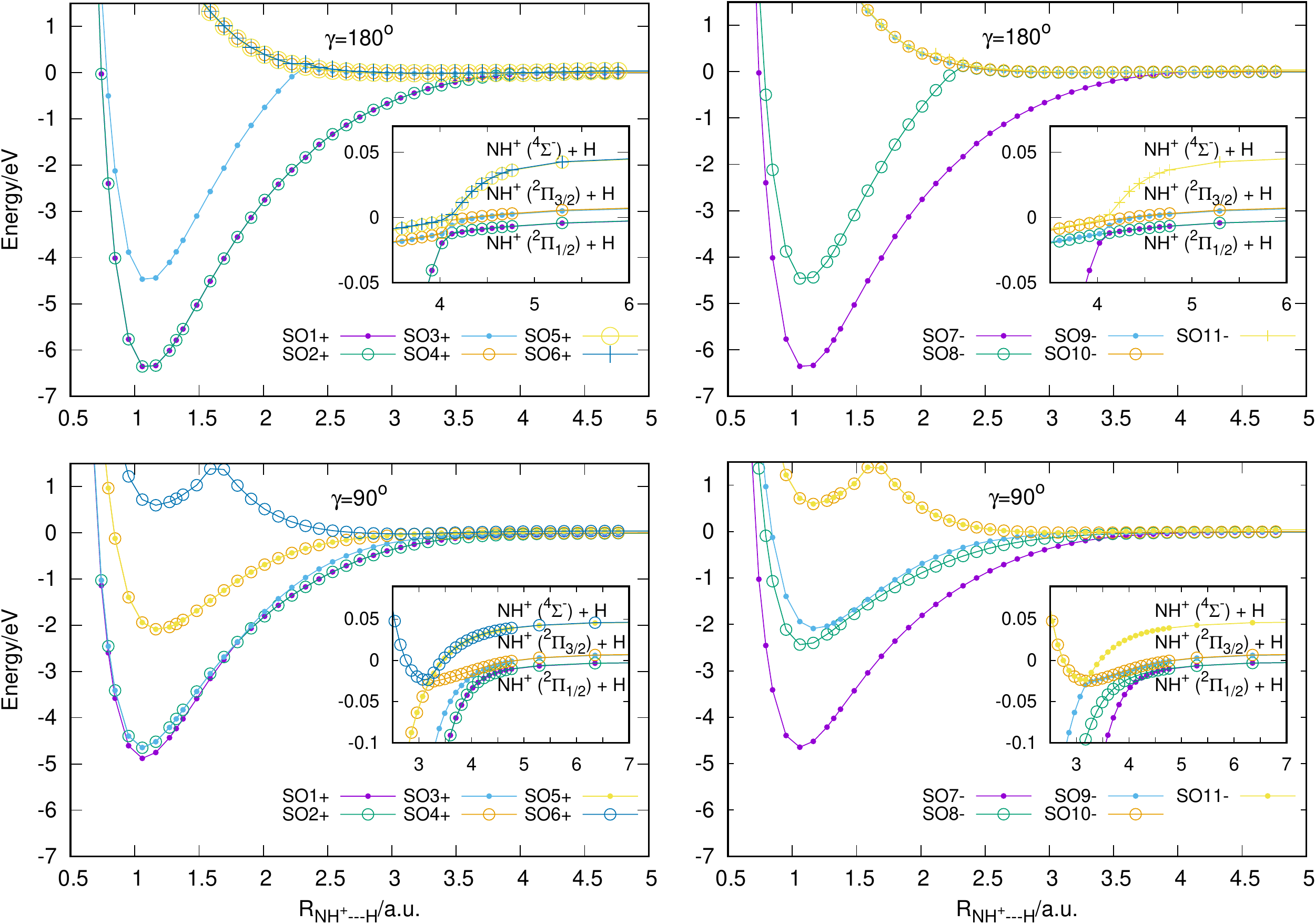}}

 \caption
 {\label{products-so-adiabatic-pes-1d}{\it Energy profiles for the 11 spin-orbit electronic states (6 even, left panels,
     and 5 odd, right panels, with respect to the reflection through the plane of the molecules) correlating
     with NH$^+$($X^2\Pi,a^4\Sigma^-$)+H($^2S$) products as a function of the R Jacobi coordinate
     for $r$= 2 a.u. and $\gamma$=90 (bottom panels) and 180 (top panels) degrees. Energies are in eV.  }}
\end{center}
\end{figure*}
\noindent

Looking at the products side in Fig. \ref{fragments}, the three lowest channels that could be energetically accessible
at the collision energies used in this work correlate with $NH^{+} (X^2\Pi) + H(^2S)$,
$NH^{+} (a^4\Sigma^{-}) + H(^2S)$ and $NH (^3\Sigma^{-}) + H^+$ asymptotes.
The $NH^{+} (A^2\Sigma^-) + H(^2S)$ and $NH^{+} (b^4\Pi) + H(^2S)$ channels are too high in energy.  

As done for reactants, a SA-CASSCF/MRCI method has been used to calculate the product channels.
The electronic states correlating with the three lowest channels without taking into account spin-orbit coupling
(in $C_{\infty v}$ point group of symmetry) are shown in Table \ref{tabla1}. 
\renewcommand{\arraystretch}{1.3}
\begin{table}[t]
\begin{center}
\begin{tabular}{l c c}
\hline
\it{Product \ asymptote} & \hspace*{2.4cm} & $C_{\infty v}$ \\
\hline   
NH$^{+} (X^2\Pi)$ + H($^2S$) &  &  $^3\Pi, ^1\Pi$ \\
NH$^{+} (a^4\Sigma^{-})$ + H($^2S$) &  &  $^5\Sigma^{-}, ^3\Sigma^{-}$\\
NH ($X^3\Sigma^{-}$) + H$^+$ & &  $^3\Sigma^{-}$  \\
\hline
\end{tabular}
\vspace*{0.2cm}
\caption{\label{tabla1}{\it Electronic states correlating with the three lowest product channels. }}
\end{center}
\end{table}

Since the ab initio calculations are done in the C$_s$ symmetry,
the state average CASSCF wavefunction has included
one $^3A'$, three $^3A''$, one $^1A'$, one $^1A''$ and one $^5A''$ states.

In the products channel, we shall use the adiabatic spin-orbit ab initio states, without considering the couplings among them,
in contrast with the treatment described above for the reactants channel.
We have focused on the states correlating with the two lowest channels,
i.e., $NH^{+} (X^2\Pi) + H(^2S)$ and $NH^{+} (^4\Sigma^{-}) + H(^2S)$ (see Table \ref{tabla1}).
That involves a total of 16 SO states. However, since we need to know the symmetry of the spin-orbit states
under the reflection respect to the molecular plane, $A'$ or $A''$,
the quintuplet electronic states have not been included because they are repulsive and
the symmetry treatment
is not yet implemented in the Molpro 2015 program. In any case,
we have checked that their omission does not affect much the accuracy of the calculations.
So, finally, 11 adiabatic spin-orbit energies have been obtained,
which are shown in Figs.~\ref{products-so-adiabatic-pes-1d}
(see also SI). Among those 11 states,
8 of them correlate with the lowest product channel, $NH^{+} (X^2\Pi_{{3/2},{1/2}})  + H(^2S)$,
and the other 3 connect  with the $NH^{+} (a^4\Sigma^{-}) + H(^2S)$ asymptote.
Fig \ref{products-so-adiabatic-pes-1d} shows the energy profiles of the 11 SO-states as a function of the R product Jacobi coordinate,
for even and odd symmetries with respect to reflection through the plane of the molecule. These curves show several crossings
among the spin-orbit states, which do not occur for all the angles. The anisotropy of the potential depend a lot on the existence or not
of such crossings. Thus, the lowest spin orbit states on each symmetry are clearly connected to the deep insertion well for $\gamma > 60^o$.
However, those intermediates presenting a crossing about $\gamma=90^o$,  present a narrower well only in the $60<\gamma<120^o$,
and this will reduce the capture probabilities, as discussed below. Finally, the higher states do not present wells and they
will be neglected in the statistical calculations presented in this work. 

Another issue which is not yet clear for this system, it is the ergicity of this reaction.
Experimentally this reaction has been found to be endoergic by 18$\pm$2 meV \cite{Marquette-etal:88}. Gerlich\cite{Gerlich:89}
compared the measured temperature dependencies on the rate constants with a statistical theory for n-H$_2$ and proposed an endoergicity of
17 meV. 
Our calculations yield an endoergicity of 80 meV, including zero point energies of reactants and products.
Below, we shall use the value of 17 meV.

\section{\label{section-quantum-statistical} Quantum statistical calculations}

The thermal reaction rate constant is defined as
\begin{eqnarray}\label{themal-rate} 
  K(T) &=& \sum_{q,\beta=1} w_{q,\beta=1}(T)\,\sum_{q' \beta' } K_{\beta q,\beta' q'}(T) \\
  {\rm with}&&
  w_{q, \beta }={ (2I_{bc}+1)(2j_{bc}+1)(2J_A+1) e^{-E_{\beta q}/k_BT} \over \sum_{q'' \beta''} (2I_{bc}+1)(2j_{bc}+1)(2J_A+1) e^{-E_{\beta'' q''}/k_BT} }\nonumber
\end{eqnarray}
where the sum is over all vibrational, rotational and electronic states of the reactants, H$_2$($X^1\Sigma^+$, v j) + N$^+$($^3P_{J_A}$),
of energy $E_{\beta q}$. In these expressions,  $\beta,q,m $ are collective quantum numbers
specifying the particular state of reactants and products. $\beta=1,2,3$ denote the arrangement channel,
H$_2$+N$^+$ ,and the two equivalent H + HN$^+$ and NH$^+$+H channels of products,
respectively.
$q=e_{bc},v_{bc}, j_{bc}, I_{bc},J_A$ are the electronic, vibrational, rotational and nuclear spin quantum numbers of the BC fragment
($I_{bc}$ =0 and 1 for para/ortho H$_2$),
while $J_A$ denotes the electronic angular momentum of the atomic fragment. Finally, $m=\Omega_{bc},\Omega_A$ are the projections
of the angular momentum of the diatomic and atomic fragment in the body-fixed z-axis, respectively, in each rearrangement channel.
$ K_{\beta=1 q,\beta' q'}(T)$ are the state-to-state rate constants, which correspond to
the Boltzmann average over the translation energy, $E$, of the reaction state-to-state cross section
\begin{eqnarray}\label{s2s-rate} 
  K_{\beta q,\beta' q'}(T) = \sqrt{ {8\over \pi \mu (k_BT)^3}}  \int dE\, E\, \sigma_{\beta q, \beta' q'}(E) e^{-E/k_BT}.
\end{eqnarray}

The cross section is obtained under the partial wave summation over
the total angular momentum, $J$, and parity under inversion of spatial coordinates, $p$, as
\begin{eqnarray}\label{cross-section}
  \sigma_{\beta q,\beta' q'}(E) &=& {\pi \over (2j_{bc}+1) (2J_A+1)\,k^2_{\beta q}(E)} \nonumber\\
   &\times& \sum_{J p}\sum_{mm'} (2J+1)\,
  P^{Jp}_{\beta q m;\beta' q' m' } (E),
\end{eqnarray}
where $k_{\beta q}= \sqrt{2\mu (E-E_{\beta q})}/\hbar$ (with $\mu$ being the H$_2$ + N$^+$ reduced mass), 
and $E$ being the total energy.

$P^{Jp}_{\beta q m;\beta' q' m' } (E)=\vert S^{Jp}_{\beta q m;\beta' q' m'}\vert^2 (E) $
are the state-to-state reaction probability from a particular initial state $(\beta,q,m)$  of
the reactants  to a final state of products ($\beta' q',m'$). This quantity can be calculated
with different methods, exact and approximate, quantum and classical. In the statistical approach \cite{Pechukas-etal:66,Miller:70}
the state-to-state reaction probability is calculated as
\begin{eqnarray}\label{statistical-reaction-probability}
  P^{Jp}_{\beta q m;\beta' q' m' } (E) = { C^{Jp}_{\beta q m }(E)  B^{Jp}_{\beta' q' m'}(E)} 
\end{eqnarray}
with the branching ratio matrix, $B^{Jp}_{\beta' q' m'}$, 
 being defined as
\begin{eqnarray}\label{branching-matrix}
 B^{Jp}_{\beta' q' m'}(E)={C^{Jp}_{\beta' q' m'}(E)
         \over \sum_{\beta'' q''  m''}  C^{Jp}_{\beta'' q'' m''}(E) }
\end{eqnarray}
where the sum in the denominator runs over all the accessible states of reactants and products.
For $J=j =0$ there are many forbidden channels. This factorization, allows to define
a capture cross section as
\begin{eqnarray}\label{capture-sigma}
  \sigma^C_{q,\beta=1} &=& {\pi  k_{vjJ_A}^{-2}(E)\over (2j+1) (2J_A+1)}
  \sum_{J p} \sum_{m} (2J+1)\quad C^{Jp}_{\beta q m}(E).
  \nonumber\\
  &&
\end{eqnarray}
In the case of very exothermic reactions, the capture cross section coincides with
the reactive cross section. In this factorization, we could define approximately
the cross section as
\begin{eqnarray}\label{complex-forming-cross-section}
  \sigma_{\beta q,\beta' q'}(E) &\approx&   \sigma^C_{\beta q}(E) \times {\cal B}_{\beta' q'}(E)
\end{eqnarray}
with
\begin{eqnarray}\label{sum-branching}
  {\cal B}_{\beta' q'}(E) = \sum_{m'} \sum_{Jp}  B^{Jp}_{\beta' q' m'}(E),
\end{eqnarray}  
which would only be accurate when the individual $B^{Jp}_{\beta' q' m'}(E)$ do not
strongly depend on $J$ and $p$. Otherwise it can only be taken as an approximation
for complex forming reactions.

The different statistical approaches depend on the procedure followed to calculate
the $C^{Jp}_{\beta q m}$ capture probabilities. In the present work we
use the quantum statistical  \cite{Rackham-etal:01,Rackham-etal:03,Alexander-etal:04,Gonzalez-Lezana:07},
and the adiabatic statistical \cite{Quack-Troe:74,Quack-Troe:75,Troe:87} approaches.

In the quantum statistical approximation,
a set of inelastic close-coupled equations is solved for each rearrangement channel independently
imposing complex boundary conditions at short distances as described in the
Appendix \ref{appendix-zticc}. For doing so, we have developed here a new program based
in the Renormalized Numerov method (called aZticc), 
as described in the Appendix. The original coupled nuclear-electronic
diabatic basis set used for  N$^+(^3P_{J_A})$+ H$_2$ reactants
is that of Ref.\cite{Jouvet-Beswick:87}, $\vert JM\Omega j J_A\Omega_Ap\rangle$, which are linear combinations
of functions
\begin{eqnarray}\label{total-basis-set}
  \vert JM\Omega j J_A\Omega_A\rangle = \sqrt{ {2J+1\over 8\pi^2}} D^{J*}_{M\Omega}(\phi,\theta,\chi)\,
  Y_{j\Omega-\Omega_A}(\gamma,0)\, \vert J_A\Omega_A\rangle,\nonumber\\
\end{eqnarray}
with parity $p=\pm$1 with respect to inversion of spatial coordinates. The treatment
is described in the SI for completeness,  where the
matrix elements of the different terms of the Hamiltonian are also shown. For products,
described in an adiabatic spin-orbit approximation, no correlation among electronic and nuclear angular
momenta is considered, and we treat them as a particular case  with
$L=S=0$.

The adiabatic statistical approach \cite{Quack-Troe:74,Quack-Troe:75,Troe:87}
uses a classical approach for the capture probability, $i.e.$
\begin{eqnarray}\label{classical-capture-probability}
  C^{Jp}_{\beta q m} =\left\lbrace
  \begin{array}{ccc}
    1  & {\rm when } & E > E_b\\
    0  & {\rm when } & E < E_b
  \end{array}
  \right. ,
\end{eqnarray}
where $E_b$ is the energy at the top of the barrier associated to the corresponding
adiabatic eigenvalue of the matrix $\overline{V}(R)$  appearing in the close-coupling equations,
in Eq.~(\ref{close-coupling-equations}).

\section{\label{results} Results and discussions}

\subsection{Quantum versus classical capture probabilities}
\begin{figure}[t]
\begin{center}
 \includegraphics[width=8.cm]{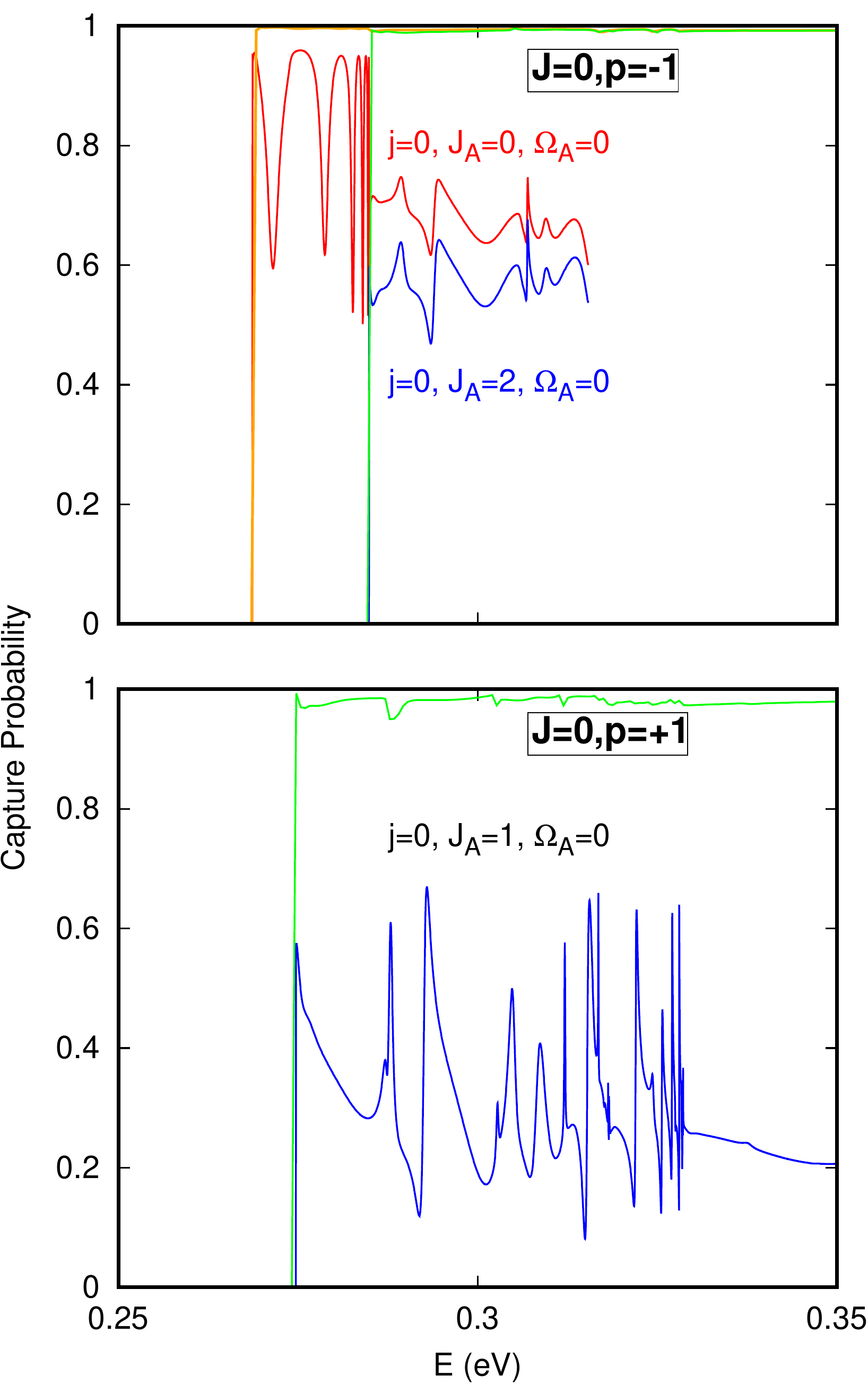}
  
 \caption{\label{capture-conSO}{\it Quantum capture probabilities obtained for the N$^+$($^3P_{J_A}, J_A=0,1,2$)
     +H$_2$(v=0,j=0) for $J$= 0, and $p=+1$, bottom panel, and $p=-1$, top panel. 
 Blue and red probabilities correspond to $R_c$ = 3 bohr, while green and orange to  $R_c$ = 4 bohr  }}
\end{center}
\end{figure}

We start by showing the quantum capture probabilities calculated with the aZticc program,
described in the appendix. The details of the numerical calculations are described in the
Supplementary Information. In Fig.~\ref{capture-conSO} the capture probabilities
obtained 
for $j$=0, $J$=0 and $p=\pm 1$ are shown, where the full  spin-orbit fine structure of N$^+$($^3P_{J_A}$) is considered
with the symmetry restrictions introduced by the treatment of Jouvet and Beswick\cite{Jouvet-Beswick:87} (in the SI).
The diabatic channel $J_A=0,\Omega_A= 0$
and $J_A=2,\Omega_A=0$  (appearing for $p = -1$)
are directly connected to the insertion well and they show a larger capture probability.
The diabatic channel 
$J_A=1,\Omega_A= 0$  presents a barrier, but it presents a non-negligible capture probability,
and this is only possible to non-adiabatic transition.

It is important to note that the quantum capture probabilities are rather different from the classical
ones, which are 1 above the barrier.
These results are obtained with a capture radius of $R_c$= 3 and 4 bohr,
as indicated in the caption of Fig.~\ref{capture-conSO}.
The captures probabilities depend on the $R_c$\cite{Gonzalez-Lezana-etal:05,Gonzalez-Lezana:07}.
To consider other $R_c$,  we can not do it by simply setting $R_{min}$= $R_c$, because the repulsive electronic states
are still open for some energies. Instead, 
we set the adiabatic-to-adiabatic transfer matrix $\overline{S_{ij}} = 1$    below $R_c$,
in Eq.~(\ref{transfer-matrices}).
By setting $R_c$=4 bohr, the quantum capture probabilities increase a lot, becoming very close to 1, $i.e.$
very similar to the classical capture probabilities using in the adiabatic statistical approach.
This demonstrate that capture probabilities
decreases because of the transition among different channels, which reflect back part of
the incoming flux. The quantum capture converges rapidly, and for $R_c$= 3.5 bohr the results
are nearly indistinguishable to those shown for $R_c$= 3 bohr in Fig.~\ref{capture-conSO}.

It is important to note here that, depending on the parity, $p$, and total angular momentum, $J$,
not all the spin orbit-states of the atom, $J_A, \Omega_A$, exist due to symmetry restrictions.
This is particularly important for $j$=0, for which only $\Omega-\Omega_A$= 0 exists. For $J=0,p=+1$, $j$=0,
in the top panel of Fig.~\ref{capture-conSO}, only the functions with $J_A=0,\Omega_A=0$ and $J_A=2,\Omega_A=0$
exist, while  for $J=0,p=-1$, $j$=0, only $J_A=1,\Omega_A=0$ appears. As $J$ and $j$ increases, and thefore
$\Omega$, more $J_A,\Omega_A$ states participate. This makes appear contributions from the three
values of the $J_A=$ 0, 1, and 2 to the reactive cross section. This occurs in the quantum as well as in the
pure adiabatic statistical approaches, as a consequence of using a coupled basis set for electronic and nuclear
angular momenta. This is not the case of previous treatments \cite{Grozdanov-etal:15,Grozdanov-etal:16},
where it is assumed that only the three lower adiabatic spin-orbit states of N$^+(^3P_{J_A})$, correlating
to $J_A= 0 $ and 1 react, while the six higher adiabatic spin-orbit states do not react.

\begin{figure}[t]
\begin{center}
 \includegraphics[width=9.cm]{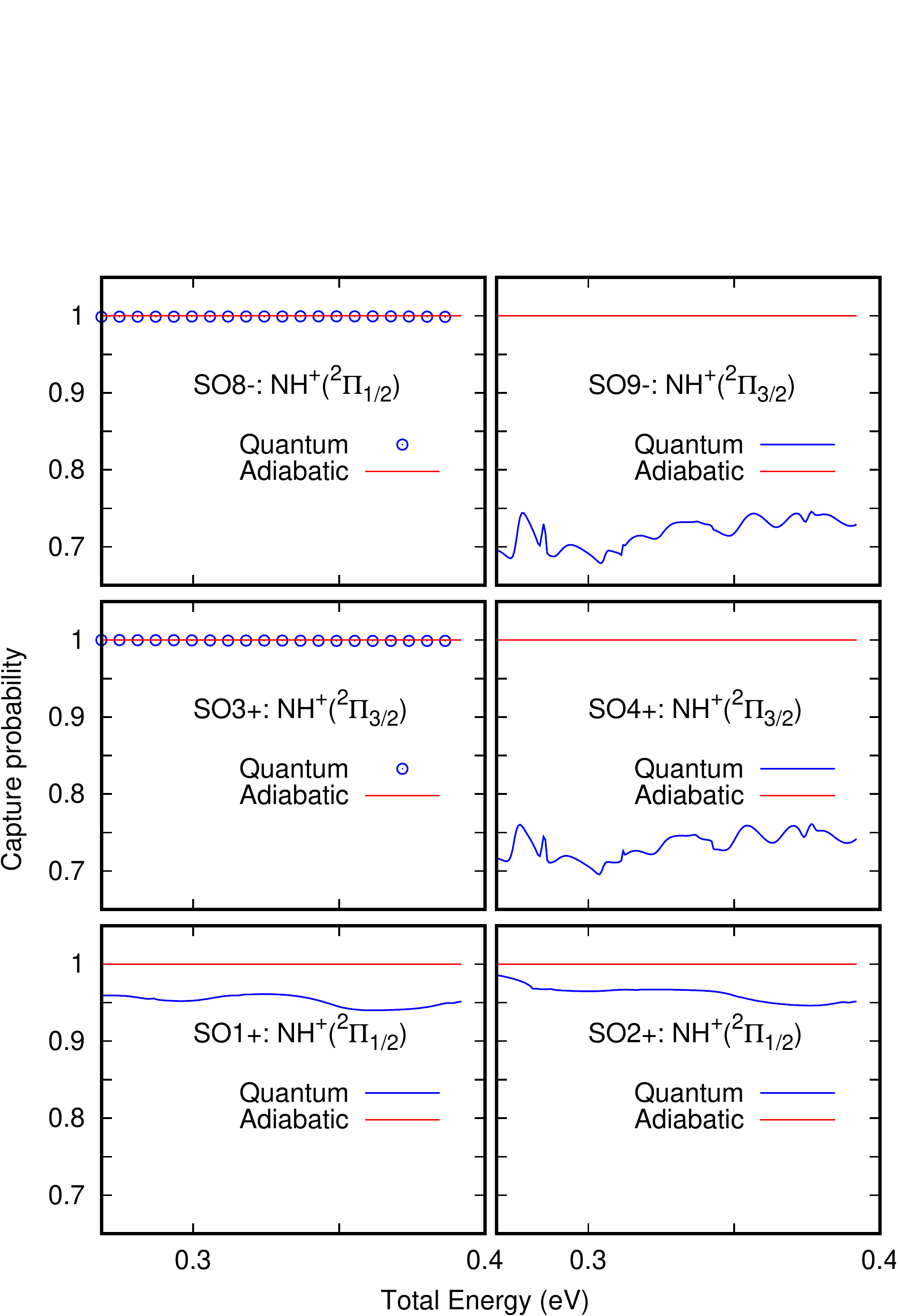}
  
 \caption{\label{capture-products}{\it Capture probabilities obtained for the NH$^+$($e,v=0,j=0$)
     +H for $J$= 0, and $p=+1$, with $R_c$= 3 bohr. Those corresponding to SO7- is equal to SO2+, SO5+ is SO4+, and SO6+, SO10- and SO11- are
     repulsive with no capture. 
  }}
\end{center}
\end{figure}

In the products channel describing the NH$^+$($ ^{2S+1}\Lambda_\Omega$) + H collision,
independent adiabatic spin-orbit states are considered in this work. The capture probabilities
calculated with the quantum and adiabatic (or classical) approaches are presented in Fig.~\ref{capture-products}.
In general the capture probabilities for a single adiabatic state are larger and with less
structure. Narrow resonances are in general absent. For states 3 and 8, the quantum capture is nearly
1 and constant, as in the adiabatic case. This is an indication that the PES anisotropy and anharmonicity
do not change from NH$^+$ products along the channel up to capture. For states 1 and 2, the quantum probability
oscillates slightly around 0.95, i.e. is rather constant, and the error of the adiabatic capture is of the order
of 5\%.  The most extreme cases are states 4 and 9, for which the quantum capture
probability is in the interval 0.7-0.75, so that we consider that in these cases
the adiabatic capture produces a relatively large error, of $\approx$ 30\%, but
nearly constant with energy. This trends persist for higher $J$, and one possible
approximation could be to multiply the adiabatic capture probability by a correction factor,
depending on the electronic state and energy independent, and this is done below for the mixed
quantum-adiabatic statistical approach.

All these results demonstrate that 
quantum capture probabilities are in general  lower than the classical ones,
which take a value of 1 for all the adiabatic electronic states.
 This
reduction is particularly important when  several electronic states are considered, for which
electronic transitions occur specially at the crossings.
When only one electronic state is considered, as it is the case for product arrangement,
the curves associated to different
channels are nearly parallel, what
reduces considerably the transitions among them before being captured. 
In these cases, the quantum capture probabilities are much closer to one, in general,
close to the adiabatic statistical approximation. 
 It should be noted,
that the anisotropy of the single adiabatic potential (see SO4+ and SO9- in Fig.~\ref{capture-products})
introduces crossings among rotational channels that can also reduce the capture.


\subsection{Total reactive cross section}

\begin{figure}[t]
\begin{center}
  
 \includegraphics[width=9.cm]{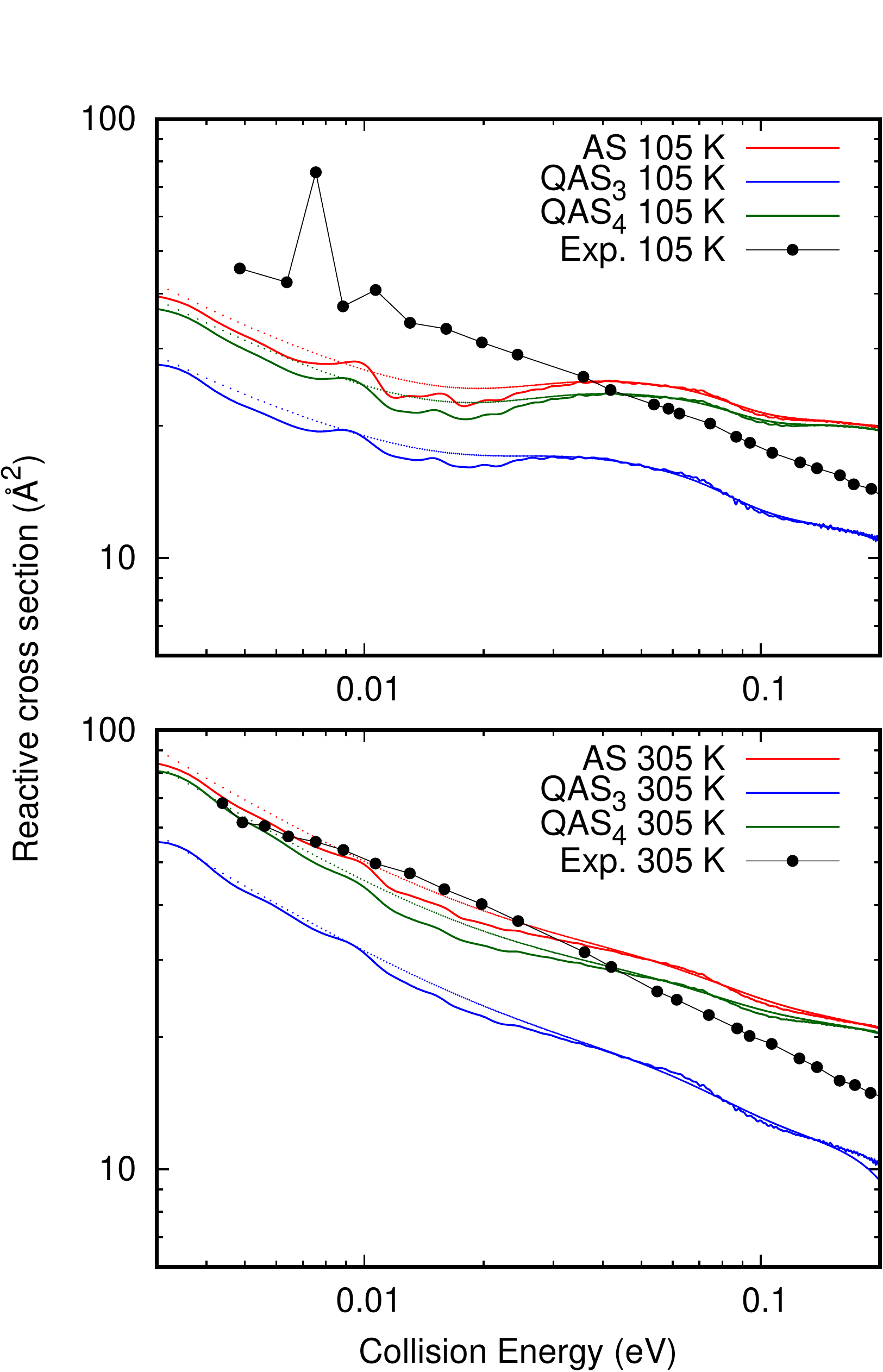}
 \caption{\label{cross-section-experiment}{\it Thermal Reactive cross section for H$_2$+N$^+$($^3P_{J_A}$),
     including the 
      thermalized at 305 K (bottom panel) and 105 K (top panel),
     obtained with the Adiabatic Statistical (AS) and mixed- Quantum/adiabatic statistical (QAS)
     methods. The experimental values are from Ref.~\cite{Sunderlin-Armentrout:94}.
     The results obtained with the two approaches are also convoluted with the Doppler broadening,
     according to Chantry\cite{Chantry:71} shown with dotted lines.
  }}
\end{center}
\end{figure}

The reaction cross section for this reaction was measured
by Sunderlin and Armentrout \cite{Sunderlin-Armentrout:94} in a rather
broad collision energy interval. In these experiments, the H$_2$ reactants
are considered at two temperatures 105 and 305 K, and the results are
broadened by the ion energy spread and Doppler broadening
\cite{Sunderlin-Armentrout:94}. In Fig.~\ref{cross-section-experiment}, the experimental
results at 305 K and 105 K are compared with those obtained in this work with the AS and the QAS
methods. The theoretical results convoluted with a gaussian accounting for the Doppler
broadening according with the method of Chantry\cite{Chantry:71} are also shown in the figure,
showing a slight increase of the cross section. However, this increase is not enough
to match the experimental results.

The QAS results, with $R_c =$ 3 and 4 bohr (QAS$_3$ and QAS$_4$, respectively),
are always below the AS results, because the quantum capture probabilities
are lower than one, as described above. At collision energies below 0.03 eV,
the AS results at 305K match very well with the experimental results\cite{Sunderlin-Armentrout:94}. This is not the case for 105K.
Above 0.03 eV, however, the AS and QAS$_4$ results are above the experimental results, while the QAS$_3$ are below.
In fact, AS/QAS$_3$ cross section difference increases with energy, because quantum captures continue decreasing, while
adiabatic captures are always one above the barrier. Above 0.03 eV (for both temperatures), the experimental
results are in between the AS and QAS$_3$ results, being the QAS$_4$ probably the best matching
the experimental results. At 0.2 eV and below, the main contributions
arise from SO1+, SO2+ and SO7-, while the other contributions are minor. The contribution
of the more excited states is relatively  small at these energies, and even if only the SO1+, SO2+ and SO7-
are included, the cross section at 0.2 eV obtained with the AS and QAS$_4$ methods are always
slightly larger than the experimental measurements. However, the QAS$_3$ is below in all the energy interval
considered here.

{     The AS treatment considers that all the flux overpassing the effective barrier is trapped, and therefore
     is treated statistically. However, when considering a quantum capture approach, we have demonstrated
     that it strongly depends on the capture radius \cite{Gonzalez-Lezana-etal:05,Gonzalez-Lezana:07}. The problem is
     therefore to determine the trapping region, without introducing artificial bias among different channels. In fact,
     considering too short capture radius includes inelastic transitions in the so-called trapping region, but only within
     the same rearrangement channel, while in the pure statistical spirit it should be considered among all rearragement channels.
     To avoid this bias, here we used the AS results as a benchmark to determine the best capture radious, without including
     any unbalance among the different rearrangement channels, what leads to the optimal value of $R_c$=4 bohr in this case, close to the average possition of the
     effective barrier used in the AS method.
    }

    It is worth mentioning, that AS and QAS$_4$ results above 0.2 eV also overestimate the reaction cross section.
The reason for this is attributed to the large mass mismatch between N$^+$ and H$_2$ subsystems, which reduces
the energy transfer probability. Statistical asumption, however, implies that energy is completely redistributted
among  all degrees of freedom, yielding to an overstimation of the reaction cross section. This is demonstrated
in the SI, where statistical results are compared with complete quantum calculations
performed with the wave packet code MADWAVE3\cite{Zanchet-etal:09b,MADWAVE3:21}
using the single adiabatic potential energy surface, PES IV of Ref.~\cite{Wilhelmsson-Nyman:92b}.

{
      The simulated cross sections change a lot varying the temperature  from
      105 to 305 K. The temperature mainly affects the rotational distribution of H$_2$
      in the cell. The cross sections for the individual initial states of the reactants
      show that H$_2$(j=0) is closed for $J_A=$ 0 and 1 below 0.01 eV,
      while it is open for all J$_A$ and for H$_2$(j=1) at all
     collision energies. This clearly explains why theoretical thermal cross section varies so much
     from 105 to 305 K. These changes, however, are not so important in the experimental
     results, which show a good agreement at 305 K with the AS and QAS$_4$ results,
     while the agreement is much worse at 105 K.

     In order to improve the experimental/theoretical agreement, different
     exothermicities have been considered. This was also done by
     Grozdanov and McCarrol\cite{Grozdanov-etal:15}, who increased the endothermicity from 18.45 meV to 23.45 meV
     to reduce their cross section, which was slightly overestimated in their approach as compared  
     to the experimental thermal cross section. However, the variation of the endoergicity,
     in all cases considered in this work, yield  rate constants in considerably worse agreement
     with the available experimental
     measurements, performed in several studies with different techniques. We therefore conclude
     that the cross sections measured by Sunderlin and Armentrout \cite{Sunderlin-Armentrout:94}
     at 105 K are also affected by the ion energy spread, as discussed by these authors, which
     is not accounted for in this work because the exact conditions of those experiments are not known.
     We also conclude that the endothermicity of 17 meV is the best choice, as shown below.
   }

\begin{figure}[h]
\begin{center}
 \includegraphics[width=7.5cm]{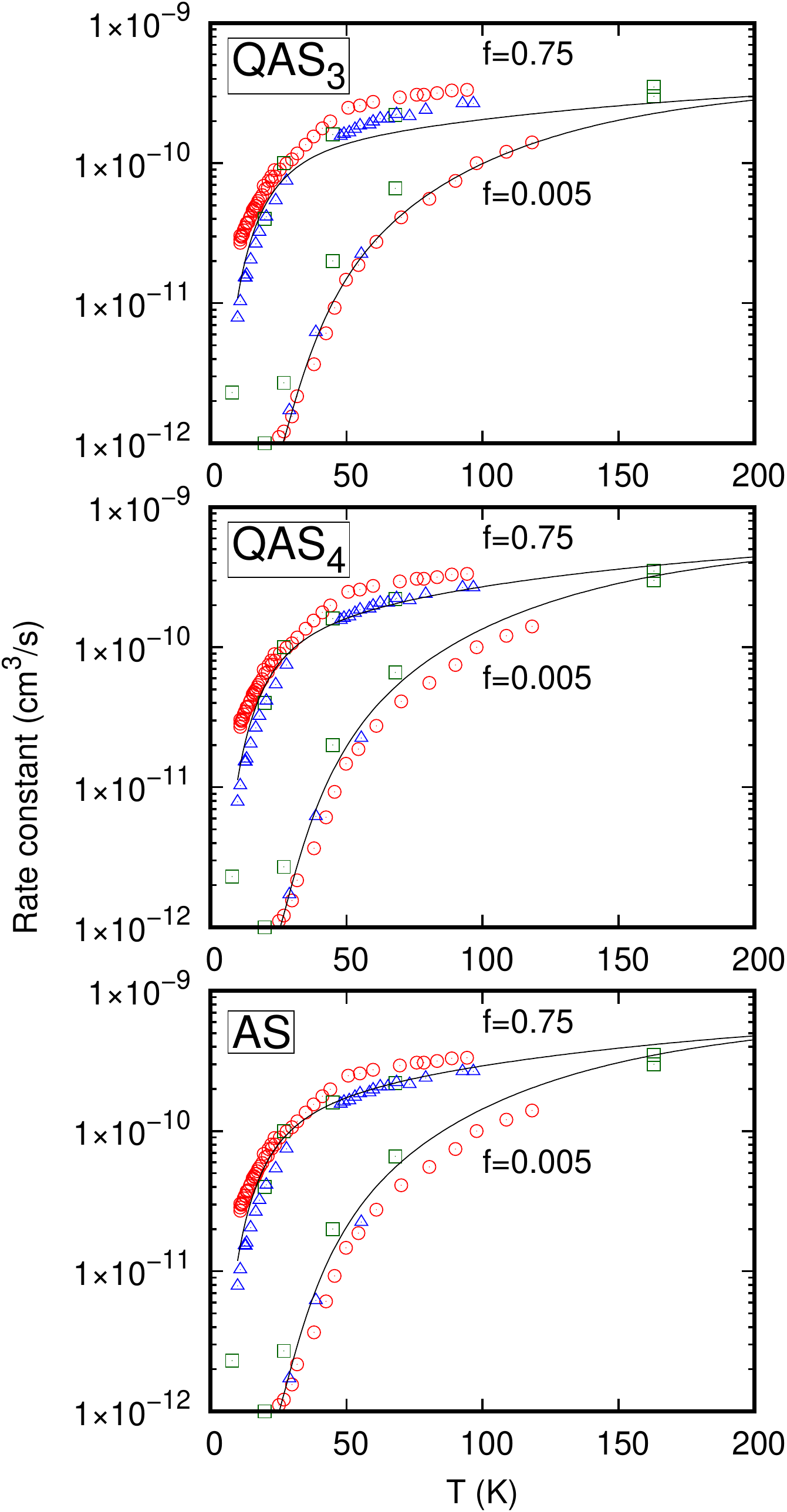}
  
 \caption{\label{rates-experiment-temperature}{\it Thermal Reactive rate constants for H$_2$+N$^+$($^3P_{J_A}$),
     obtained for the two limiting experimental o-H$_2$ fractions, f=0.005 and f=0.75 (n-H$_2$), as a funtion of temperature.
     Symbols are the experimental results: open circles by Zymak and et al.\cite{Zymak-etal:13}, open square are taken from Table II of Marquette et al.\cite{Marquette-etal:88}
     (the value list at 27 K for n-H$_2$ has been corrected to read 2.7 10$^{-12}$) and open triangles are the results of Fanghanel\cite{Fanghanel:18}.
     Lines are the simulated rate constants with QAS$_3$ (top panel), QAS$_3$ (middle panel) and the AS (bottom panel) described in this work.
  }}
\end{center}
\end{figure}

\subsection{Rate constants}

The thermal rate constants for ortho-H$_2$ fraction $f$=0.005 and 0.75, of H$_2$ are shown in Fig.~\ref{rates-experiment-temperature}
and compared with the available experimental data, for the AS (bottom panel), QAS$_4$ (middle panel) and QAS$_3$ (top panel) methods.
There is a rather good qualitative agreement between the two simulated rate constants (AS and QAS$_3$ and QAS$_4$) and
the experimental results. The QAS$_3$ results for f=0.005 agree very well with the experimental measurements of Zymak and et al.\cite{Zymak-etal:13},
and for f=0.75 lies in between the three sets of experimental results for temperatures below 50 K. However, for $T\>$ 50 K and f=0.75,
the QAS$_3$ results are considerably lower than any set of experimental results. The QAS$_4$ and AS results are in between all the sets
of experimental data in the whole temperature interval considered here, being in general closer to those of Fanghanel\cite{Fanghanel:18}.
The difference between experimental results allows to establish a certain error, probably due to the exact ortho-H$_2$ fraction f. 

\begin{figure}[h]
\begin{center}
 \includegraphics[width=8.cm]{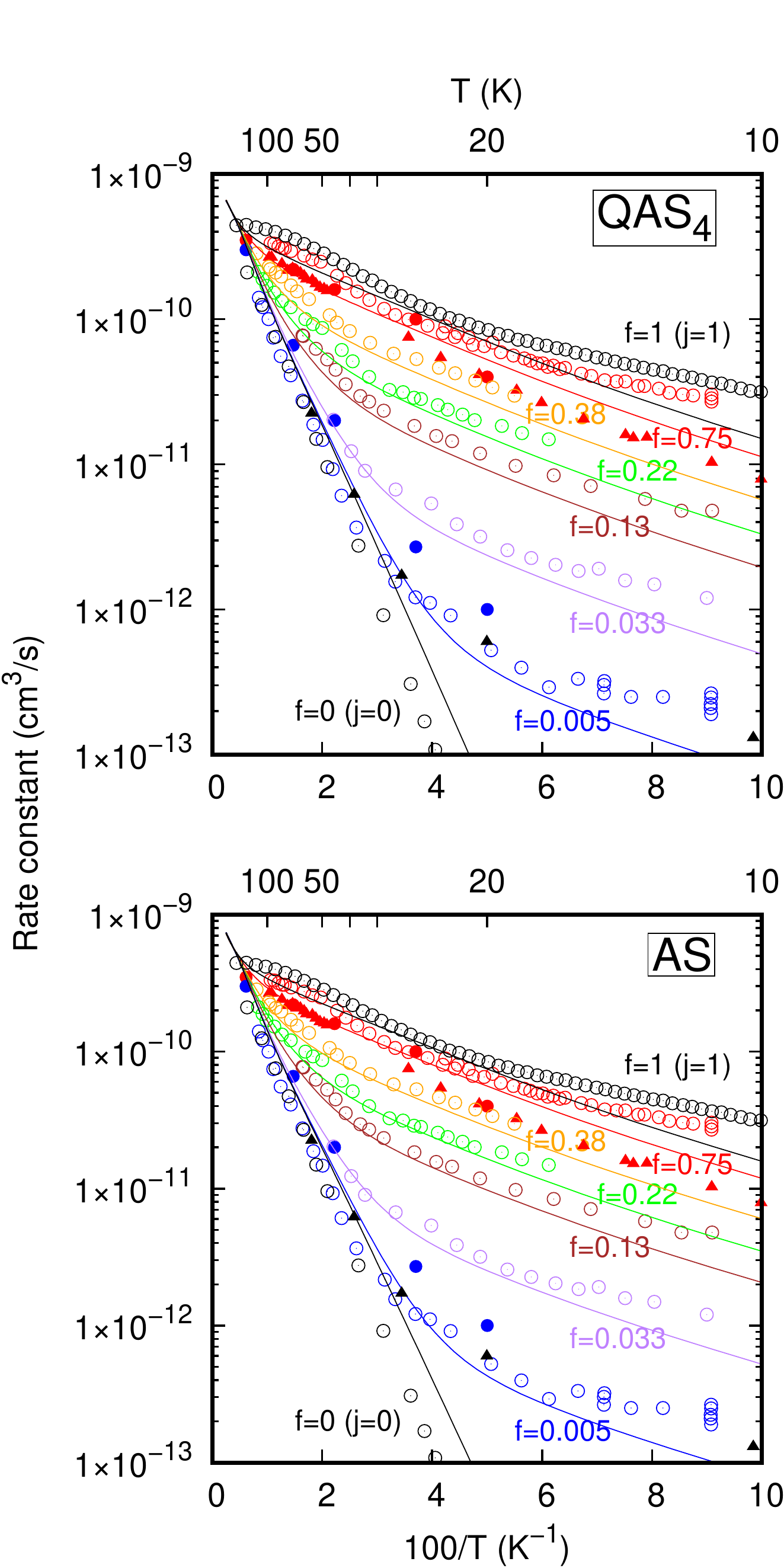}
  
 \caption{\label{rates-experiment}{\it Thermal Reactive rate constants for H$_2$+N$^+$($^3P_{J_A}$),
     obtained for different o-H$_2$ fractions, as described by Zymak and et al.\cite{Zymak-etal:13}.
     Open circles are the experimental results extracted from Fig.~4by Zymak and et al.\cite{Zymak-etal:13}
     (note that for f=0 and 1, their values
     are extrapolated). Full circles are taken from Table II of Marquette et al.\cite{Marquette-etal:88}.
     Triangles are the
     experimental results of Fanghanel\cite{Fanghanel:18}. Top panel show the mixed Quantum and
     Adiabatic statistical results for $R_c$= 4 bohr (QAS$_4$),
     bottom panel show the pure Adiabatic Statistical results (AS).
     In all cases, the population of the $J_A$ spin-orbit states correspond to a Boltzmann distribution.
  }}
\end{center}
\end{figure}

The variation of the rate constants for more values of f are shown in Fig.~\ref{rates-experiment}, for the two best theoretical
results, QAS$_4$ (top panel) and AS(bottom panel).
For $f=0$ (para-H$_2$), the experimental results of Zymak {\it et al}\cite{Zymak-etal:13} (which
were extrapolated) are in better agreement with the QAS results than with the pure AS.  However, for $f=1$ (ortho-H$_2$)
the agreement at higher temperatures is better for the AS results than for the QAS. This is probably because the
AS results are larger at 0.01 eV than the QAS, and in better agreement with the experimental cross sections,
in Fig.~\ref{cross-section-experiment}. The overall agreement of the two simulations, AS and QAS, is in general
excellent for temperatures between 20 and 100 K, and the increase of the error for T$< 20$ K could be attributted to
small contamination of ortho/para ratios, as well as to inaccuracies of the simulations.

The agreement between the two sets of experimental rate constants also show some discrepancies. These discrepancies
are similar in magnitude to that between simulations and experiments. It is important to note the large
variation of the rate constant as a function of the ortho-H$_2$ fraction, $f$, due to the fact that the reaction
is exothermic for ortho-H$_2$($j$=1), while it is closed for para-H$_2$($j$=0), whose ratios may change slightly.

\begin{figure}[h]
\begin{center}
 \includegraphics[width=9.cm]{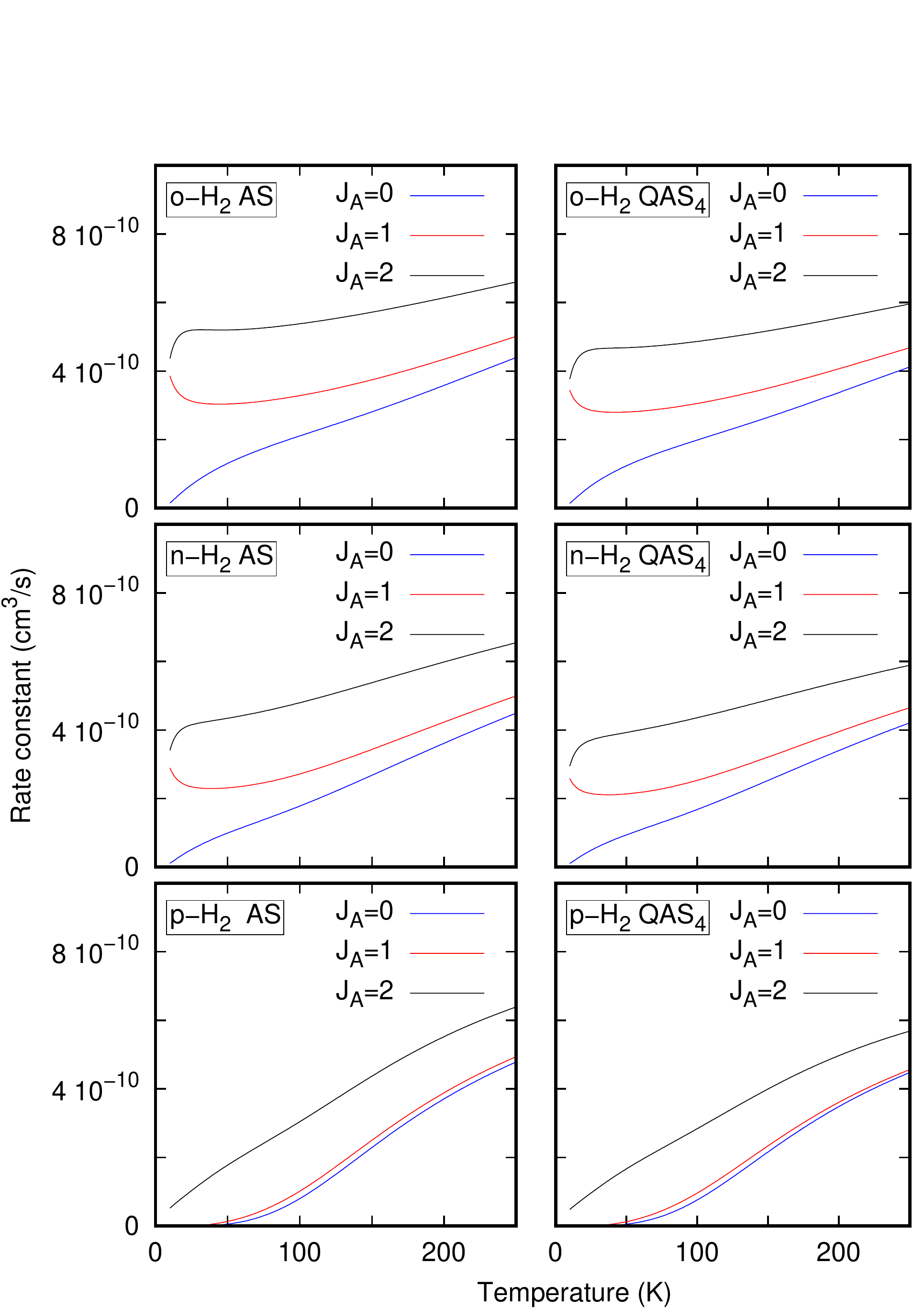}
  
 \caption{\label{rates-JA}{\it Thermal Reactive rate constants for H$_2$+N$^+$($^3P_{J_A}$),
     for ortho, natural and para H$_2$ for each  N$^+$($^3P_{J_A})$ spin-orbit state, obtained
     with the AS (left panels) and QAS$_4$ (right panels) methods.
  }}
\end{center}
\end{figure}
Moreover, a similar situation holds for the spin-orbit states of N$^+(^3P_{J_A})$: for H$_2$($j$=1) all $J_A$=0 and 1
states are open, while for H$_2$($j$=0) only $J_A$=2 is open. The individual rate constants for each $J_A$ spin-orbit
state and different ortho-H$_2$ fractions are shown in Fig.~\ref{rates-JA}. In the two formalisms, AS and QAS,
the rate constants for the 3 spin-orbit states are non-zero. Such situation may introduce changes
in the experimental determinations of the state specific rate constants, as discussed
by Zymak {\it et al.}\cite{Zymak-etal:13} and Fanghanel\cite{Fanghanel:18}.

AS and QAS methods yields to rather different rate constants for each individual $J_A$ spin-orbit state.
The AS method tends
to produce a progression $J_A$= 0, 1 and 2, with the rate for $J_A$=2 being the larger, simply
because it correspond to the most endothermic case. The situation varies a lot for the QAS results,
for which the rates for all $J_A$ are closer and their relative importance varies with temperature.
This result is  a consequence of the explicit treatment of transitions among spin-orbit states, using  correlated
electronic-nuclear diabatic basis set . Since
this is accounted for more exactly in the QAS method, in contrast to the AS one, we conclude that
the QAS $J_A$-dependent rate constants are more accurate. The numerical values of the $J_A$-dependent rate constants
are given in the SI. Our results are in general in
better agreemnet with the experimental results of Fanghanel\cite{Fanghanel:18}, where the reactivity
of N$^+(^3P_{J_A=2})$ is considered to be non-zero, as it is demonstrated in this work.

  { The accurate determination of the reaction  rate constants  is important to improve the accuracy of
    astrophysical models. 
  The rate constant available in the Kida Data base for this reaction at low temperatures
  corresponds to the value reported by Marquette et al.\cite{Marquette-etal:88} for the n-H$_2$ (corresponding to
  an ortho-fraction of f=0.75). These experimental values are compared with the present results in Figs.~ 8 and 9.
  This reaction, however, strongly depends on the initial rotational
  state of H$_2$ and also on the spin-orbit state of N$^+(^3P_{J_A})$ , as shown in this work. In detailed astronomical models,
  it is important to incorporate the  specific rate, at least  for ortho and para hydrogen. For this reason
  we provide in the Supplementary Information  the parameters obtained in a fit of the
  numerical rate constants obtained in this work, and shown in Fig. 10, for each ortho-fraction of H$_2$ and
  each electronic $J_A$ value for N$^+$, listed in a Table. 
}

\section{Conclusions}\label{section-conclusions}

In this work we have studied the spin-orbit dependence of the rate constants
for the N$^+(^3P_{J_A})$+ H$_2$ $\rightarrow$ H + NH$^+(^2\Pi_{1/2,3/2})$. The potential
energy surfaces on reactants and products channels have been calculated separately,
using accurate {\it ab initio} methods. In the reactants  N$^+(^3P_{J_A})$+ H$_2$ channel,
the couplings among the spin-orbit states have been calculated, using a diabatization
method together with a model based on  atomic spin-orbit localized in the N$^+$ cation.
This method has been compared with accurate {\it ab initio} calculations showing excellent
agreement. The  NH$^+(^2\Pi_{1/2,3/2},^4\Sigma^-)$+H products potential energy surfaces have been calculated
in the adiabatic spin-orbit approximation.

To account explicitly for the spin-orbit couplings, the treatment
of Jouvet and Beswick\cite{Jouvet-Beswick:87} have been implemented
within two statistical models: an adiabatic statistical (AS) model
and a mixed quantum and adiabatic statistical (QAS) method. A variation
of the renormalized Numerov method has been developed to treat open-quantum
boundary conditions, needed to calculate quantum capture probabilities, used
in the mixed quantum-adiabatic statistical method.

It is worth noting, that the AS model provide quite accurate rates for all spin-orbit
states of N$^+(^3P_{J_A})$, $J_A$= 0, 1 and 2, when the basis is formed by proper symmetry functions combining
electronic (spin and orbital) and nuclear angular momenta. On the contrary, when the adiabatic
approximation is  done at the spin-orbit electronic states alone first, only the first
3 spin orbit states (correlating to $J_A$= 0 and 1) can contribute to the reactive cross section and rate constants.

Thermal cross section and rate constants have been calculated and compared with
the available experimental measurements. The calculated  thermal rate constants
for different ortho fractions of H$_2$ show reasonable good agreement
with the experimental measurements of Marquette  {\it et al.}\cite{Marquette-etal:88},
Zymak {\it et al.}\cite{Zymak-etal:13} and Fanghanel\cite{Fanghanel:18},
confirming an endothermicity of
17 meV. We find that the three $J_A$ spin-orbit states have an appreciable
contributions rate constants for all the o-H$_2$ fractions, f, measured. In particular,
the possible effect of $J_A$=2 in the determination of the rate constants for f= 0 and 1
(not directly measured) was not taken into account by Zymak {\it et al.}\cite{Zymak-etal:13}
and it was included and discussed by  Fanghanel\cite{Fanghanel:18}. We demonstrate here, that
it is important to be included for this system, since there are many different energy thresholds,
for reactants ($J_A$ and $j$ values) and products, and are of particular interest
for astro physics models of cold molecular clouds.

\section{Supplementary Material}

See supplementary material for detailed description of the {\it ab initio} calculations for the reactants
and product channels, for the computational details of the dynamical calculations, the treatment used to
treat the collisions of open shell atoms with closed shell diatomic molecules, and the state-specific rate constants
for the different spin-orbit states of N$^+(^3P_{J_A})$ and ortho and para H$_2$ are described
and provided in separate files.\\

\section{Acknowledgements}

We want to thank Prof. P. Armentrout for providing us the experimental values
of the cross section measurements. 
 The research leading to these results has received funding from
 MICIYU under grant No. PID2021-122549NB-C2.
 The calculations have been performed in Trueno-CSIC and CCC-UAM. 

\section{Data availability}
The data that support the findings of this study are available from the authors
upon reasonable request.

\appendix

 \section{Quantum capture method\label{appendix-zticc}}

 \subsection{Diabatic representation}
 
 The method used here to evaluate the quantum capture probabilities is very similar
 to that previously described by Rackham {\it et al.}\cite{Rackham-etal:01}. Expanding the total
 wave function in a diabatic basis set as
 \begin{eqnarray}\label{close-coupling-wvf}
   \Psi^\beta(R,{\bf x}) = \sum_n \Phi^\beta_{\alpha}(R) \varphi_\alpha({\bf x})
 \end{eqnarray}
 The close-coupling
 equations can be written as
\begin{eqnarray}\label{close-coupling-equations}
 && {\partial^2 \Phi^\beta_\alpha(R) \over \partial R^2} = {2\mu\over \hbar^2} \sum_{\alpha'}
  \left\lbrace V_{\alpha\alpha'}(R) - E \delta_{\alpha\alpha'}\right\rbrace \Phi^\beta_{\alpha'}(R) \\
  &&\quad\quad\equiv
               \overline{\Phi}''(R)   =
               {2\mu\over \hbar^2} \left\lbrace \overline{V}(R) - \overline{\bf 1} E\right\rbrace  \overline{\Phi} (R)
\nonumber
\end{eqnarray}
where $\alpha,\beta$ denotes the collections of quantum numbers 
needed to specify the channels, and $\overline{\Phi}$ is a vector and $\overline{V}$ is a matrix.
Eq.(\ref{close-coupling-equations})  are solved here using a Numerov-Fox-Goodwin
or renormalized Numerov method \cite{Roncero-etal:19,Gadea-etal:97}, in which each
of the quantities is discretized in a radial grid of $N$ equidistant  $R_i$ points, with $\Delta=R_{i+1}-R_i$.
Denoting ${\boldsymbol \Phi}(R=R_i)= {\boldsymbol \Phi}_i$
and $\overline{ V}(R=R_i)= \overline{ V}_i$, and doing a Taylor expansion of the coefficients and their second
derivatives, a three points Numerov relationship is found
\begin{eqnarray}\label{Numerov-matrices}
\overline{\alpha}_{i-1} \overline{\Phi}_{i-1} \quad
+ \quad \overline{\beta}_i \overline{\Phi}_i \quad +
\quad \overline{\gamma}_{i+1} \overline{\Phi}_{i+1}\quad =\quad 0
\end{eqnarray}
where
\begin{eqnarray}
\overline{\alpha}_{i-1} &=& \overline{\bf 1}
        -{\Delta^2\over 12} \left\lbrace {2\mu\over\hbar^2}
      (\overline{V}_{i-1}-E\overline{\b 1})\right\rbrace \nonumber\\
\overline{\beta}_{i} &=&-2\overline{\bf 1} -{10\Delta^2\over 12} \left\lbrace{2\mu\over\hbar^2}
(\overline{V}_{i}-E\overline{\b 1})\right\rbrace \\
\overline{\gamma}_{i+1} &=& \overline{\bf 1} -{\Delta^2\over 12} \left\lbrace {2\mu\over\hbar^2}
(\overline{V}_{i+1}-E\overline{\bf 1})\right\rbrace \nonumber,
\end{eqnarray}
with an error proportional to $\Delta^6$.

The Fox-Goodwin algorithm consists in
defining
\begin{eqnarray}\label{time-independent-propagation-matrix}
\overline{\Phi}_i\quad =\quad \overline{\cal R}_i\quad \overline{\Phi}_{i+1}
\end{eqnarray}
so that imposing the boundary condition at $i$=1, the ${\cal R}_i$
is propagated according
\begin{eqnarray}\label{R-propagator}
\overline{\cal R}_i &=& -\bigg\lbrace
\overline{\alpha}_{i-1} \overline{\cal R}_{i-1}\quad
+\quad\overline{\beta}_i\bigg\rbrace^{-1} \quad
\overline{\gamma}_{i+1}
\nonumber
\end{eqnarray}
until $i$=N, where the second boundary conditions of incoming plus outgoing
waves are imposed.

Usually,  a real $\overline {\Phi}_i$  is propagated
to simplify the calculation because the potential is also real and
a regular solution with $\overline {\Phi}_{i=0} = 0$ is imposed
because $\overline {V}_1 > E$ for all the channels involved.

On the contrary, in the case of capture in a well,
it is assumed that  the  $\overline {V}_1 < E$
for some of the channels. In order to impose the boundary condition,
a transformation to a new adiabatic basis is first done by diagonalizing
the  potential matrix at $i$=1 as
\begin{equation}\label{adiabatic-eigenvectors}
  \overline {V}_i  \quad \overline{ ^iT} =  \overline{D}_i \quad\overline{^iT},
\end{equation}
where $\overline {D}$ is a diagonal matrix with the eigenvalues and
$\overline {T}$ are the transformation matrix. In this adiabatic
representation the coefficients are denoted $\Psi$ to be distinguished
from those of the original ``diabatic'' basis, and the boundary outgoing conditions
are applied as
\begin{eqnarray}\label{inner-boundary-condition}
  ^1\Phi_\alpha(R< R_1) =
  \left\lbrace
  \begin{array}{cc}
    0 & {\rm if}\, D_\alpha > E\\
     & \\
  -i\sqrt{ {2\mu \over \pi\hbar^2 k_\alpha}} S^T_\alpha(E) e^{-ik_\alpha R} &{\rm if}\,  D_\alpha < E
       \end{array}
   \right.
\end{eqnarray}
where it is being assumed that $\overline {V}_i= \overline {V}_1$ for $i < 1$, $i.e.$ that
the potential is constant at $R$ distances shorter than $R_1$, the capture
distance. In this expression $k_\alpha= \sqrt{2\mu (E-d_{\alpha})}/\hbar$
and  $\left\vert S^T_\alpha(E)\right\vert^2$ is the capture probability,
since it correspond to the flux going to $R< R_1$. Under this assumptions
in the adiabatic representations we have
\begin{eqnarray}\label{adiabatic-initialR}
  \overline{\cal A} = \lbrack \overline{^1\Phi}_1\rbrack^{-1} \quad \overline{^0\Phi}_0
  =\left\lbrace
  \begin{array}{cc}
    e^{-\vert k_\alpha\vert  \Delta } & {\rm if}\, D_\alpha > E\\
     & \\
  e^{-ik_\alpha \Delta } &{\rm if}\,  D_\alpha < E
       \end{array}
   \right.
\end{eqnarray}
Transforming back to the diabatic representation in which the integration
is performed, we get (i=1)
\begin{eqnarray}
\overline{\cal R}_i = \overline{^iT}^{ -1} \quad\overline{\cal A}\quad \overline{^iT}.
\end{eqnarray}

After defining the propagation matrix in the first point of the grid,
 $\overline{\cal R}_i$ is iteratively
propagated from $i$=2 to N, where the usual incoming/outgoing boundary conditions
are imposed as
\begin{eqnarray}\label{complex-outer-boundary-conditions}
  \Phi_{\alpha}(R_N) =
  \left\lbrace
  \begin{array}{cc}
   e^{-\vert k_{\alpha}\vert R}  & {\rm if}\, V_{\alpha,\alpha}(R_N) > E\\
     & \\
     i \sqrt{{2\mu\over \pi\hbar^2 k_\alpha}}
     \left\lbrack e^{-i( k_\alpha R-\ell\pi/2)}\delta_{\beta,\alpha}\right. &\\
      \left. \quad\quad\quad - S^R_{\alpha\beta}(E) e^{i(k_\alpha R-\ell\pi/2)} \right\rbrack
         & {\rm if}\,  V_{\alpha,\alpha}(R_N) < E
       \end{array}
   \right.
\end{eqnarray}
In the usual procedure,  $\overline{\cal R}_i$ is real and real boundary conditions
are imposed to calculate the symmetric reaction matrix, $K$, and from it the
S-matrix which is unitary. In the present case, 
 $\overline{\cal R}_1$ is complex, all this procedure
is done in the complex plane, and assuming that the integration is done until sufficiently
long distance, Eq.(\ref{complex-outer-boundary-conditions}) is
also fulfilled at $R_{N-1}$, and the S-matrix is directly obtained from the propagation matrix as
\begin{eqnarray}\label{S-matrix}
  \overline{\cal S}^R &=& \left\lbrack  \overline{\cal R}_{N-1} \overline{\cal P}_N  - \overline{\cal P}_{N-1}\right\rbrack^{-1} \nonumber\\
  &\times& \left\lbrack\overline{\cal M}_{N-1}-  \overline{\cal R}_{N-1} \overline{\cal M}_N \right\rbrack
\end{eqnarray}
where ${\cal M}_i$ and  ${\cal P}_i$ are diagonal matrices with elements defined as
\begin{eqnarray}\label{outer-complex-boundary-matrices}
  {\cal M}_{\alpha\alpha}(R_i) &=&
  \left\lbrace
  \begin{array}{cc}
   e^{-\vert k_{\alpha}\vert R}  & {\rm if}\, V_{\alpha,\alpha}(R_i) > E\\
     & \\
      \sqrt{{2\mu\over \pi\hbar^2 k_\alpha}}
      e^{-i(k_\alpha R-\ell\pi/2)} 
         & {\rm if}\,  V_{\alpha,\alpha}(R_N) < E
  \end{array}
  \right.
  \nonumber\\
  \\
  {\cal P}_{\alpha\alpha}(R_i) &=&
  \left\lbrace
  \begin{array}{cc}
   0  & {\rm if}\, V_{\alpha,\alpha}(R_i) > E\\
     & \\
      \sqrt{{2\mu\over \pi\hbar^2 k_\alpha}}
      e^{i(k_\alpha R-\ell\pi/2)} 
         & {\rm if}\,  V_{\alpha,\alpha}(R_N) < E
       \end{array}
  \right.
  \nonumber
\end{eqnarray}
The resulting $S^R(E)$, in Eq.(\ref{complex-outer-boundary-conditions}) is not, in general, unitary.
This is evident by inspection of Eq.(\ref{inner-boundary-condition}),
since for those channels with $D_\alpha < E$ there is a flux that is trapped at distances
$R< R_1$. If all $D_\alpha > E$ the normal situation is got, and the $S^R$-matrix becomes unitary.
The capture probability for a given initial channel is then obtained as
\begin{eqnarray}\label{capture-probability}
  C_\beta(E)= \vert S^T_{\beta\beta}\vert ^2 = 1- \sum_\alpha \vert S^R_{\beta\alpha}\vert ^2.
\end{eqnarray}

\subsection{Adiabatic-by-sectors  representation}
 
The number of channels increases very rapidly with total angular momentum, specially
with many electronic states, as considered here. In order to reduce the number of
channels we have implemented a variant of the adiabatic-by-sectors
method \cite{Johnson-Levine:72,Garret-etal:81,Lepetit-etal:86,Schwenke-etal:87}.
In brief, this method consists in diagonalizing the $ \overline{V}_i$ matrix in the close-coupling
equations, Eq.~(\ref{close-coupling-equations}) as in Eq.~(\ref{adiabatic-eigenvectors}).
The new adiabatic functions, $^iA_p$, depend on the collision coordinate R$_i$,
are  expressed in the original diabatic basis set as
\begin{eqnarray}\label{adiabatic2diabatic-transformation}
  \vert ^iA_p \rangle &=& \sum_\alpha \vert \varphi_\alpha\rangle  ^iT_{\alpha p}
  \equiv \overline{^iA} =\overline{\varphi}\quad \overline{^iT}\nonumber
  \\
  \vert \varphi_\alpha\rangle   &=& \sum_p  \vert ^iA_p \rangle ^iT_{\alpha p}
  \equiv \overline{^i\varphi} =\overline{A}\quad  \overline{^iT}^{ -1}\nonumber.
\end{eqnarray}

The total wave function, Eq.~\ref{close-coupling-wvf}, in the new basis set takes the form
\begin{eqnarray}
  \Psi^\beta(R,{\bf x}) = \sum_p \vert ^iA_p\rangle \, ^i\Phi^\beta_p
  \rightarrow  \overline{^i\Phi}  = \overline{^iT}\quad \overline{\Phi}
\end{eqnarray}.

The Numerov auxiliary matrices in Eq.~(\ref{Numerov-matrices})
can be re-expressed in the adiabatic representation as
\begin{eqnarray}\label{adiabatic-Numerov-matrices}
  \overline{^{i-1}\alpha}_{i-1} &=&\overline{ ^{i-1}T}^{ -1}\quad \overline{\alpha}_{i-1}\quad \overline{^{i-1}T}
     = \overline{\bf 1}
        -{\Delta^2\over 12} \left\lbrace {2\mu\over\hbar^2}
                                    (\overline{D}_{i-1}-E\overline{\b 1})\right\rbrace \nonumber\\
  && \\
       \overline{ ^i\alpha}_{i-1} &=&  \overline{^{i}T}^{ -1}\quad \overline{\alpha}_{i-1}\quad \overline{^{i}T}
          = \overline{S}_{i i-1}\quad  \overline{^{i-1}\alpha}_{i-1}\quad  \overline{S}^{-1}_{i i-1},\nonumber
\end{eqnarray}
and similarly for $\overline {\beta}_i$ and $\overline{\gamma}_{i+1}$ matrices, where
the transfer matrix has being defined as
\begin{eqnarray}\label{transfer-matrices}
  \overline{S}_{i j} =\overline{^iT}^{ -1}\,\, \overline{^jT}.
\end{eqnarray}

Doing some algebra, the recurrence equation of the propagation matrix, Eq.~(\ref{R-propagator})
becomes
\begin{eqnarray}\label{adiabatic-R-propagator}
\overline{^{i+1}\cal R}_i &=& -\bigg\lbrace
\overline{^{i+1}\alpha}_{i-1} \quad\overline{ ^{i+1}\cal R}_{i-1}\quad
+\quad\overline{ ^{i+1}\beta}_i\bigg\rbrace^{-1} \quad
\overline{^{i+1}\gamma}_{i+1} \nonumber\\
\end{eqnarray}
where $\overline{^{i+1}\cal R}_i$ is the propagation matrix conecting the function $i$ and $i+1$,
represented in the adiabatic basis at $i+1$, and
\begin{eqnarray}
  \overline{ ^{i+1}\cal R}_{i-1} =\overline{S}_{i+1 i} \quad \overline{^{i}\cal R}_{i-1}\quad  \overline{S}^{ -1}_{i+1 i}
\end{eqnarray}
with $\overline{^{i}\cal R}_{i-1}$ being obtained in the previous iteration.
Eq.~(\ref{adiabatic-R-propagator}) can be iteratively solved analogously to procedure in the diabatic representation
with the extra-effort of transforming the matrices from one point to the following one.
In this adiabatic representation, the first value is that $^1{\overline{\cal R}}_0=  \overline{\cal A}$,
  defined in Eq.~(\ref{adiabatic-initialR}). Also, the diabatic and adiabatic representation
  coincide for $i=N$, so that $\overline{^{N}\cal R}_{N-1}\equiv \overline{\cal R}_{N-1} $
  and the outer boundary conditions are imposed as in Eq.~(\ref{complex-outer-boundary-conditions}).

The advantage of using the adiabatic-by-sector propagation is that we can reduce
the number of channels, by keeping only those which has an energy $E_a(R) < E_{cut}$
for all distances $R$. This propagator has been implemented in the aZticc program
for A + BC collisions, for the case of open shell atom + closed diatom following
the work of Jouvet and Beswick \cite{Jouvet-Beswick:87}, described in the Supplementary
Information.

%

 \end{document}